\documentclass[12pt]{article}
\usepackage{epsfig} 
\usepackage{amsbsy}
\newcommand\as{\alpha_{\rm \scriptscriptstyle S}}
\newcommand\ecm{E_{\rm \scriptscriptstyle CM}}
\newcommand\xb{x_{\scriptscriptstyle B}}
\newcommand\xbo{x_{\scriptscriptstyle B_1}}
\newcommand\xbt{x_{\scriptscriptstyle B_2}}

\def\sigB{\sigma_{\scriptscriptstyle B}}
\def\sigb{\sigma_{b}}

\newcommand     \MSB            {\ifmmode {\overline{\rm MS}} \else
                                 $\overline{\rm MS}$\fi}
\newcommand\xhb{\hat{x}_{b}}
\newcommand\xhbo{\hat{x}_{b_1}}
\newcommand\xhbt{\hat{x}_{b_2}}
\newcommand\LambdaQCD{\Lambda_{\scriptscriptstyle \rm QCD}}
 
\def\beq{\begin{equation}} 
\def\eeq{\end{equation}} 
\def\beqn{\begin{eqnarray}} 
\def\eeqn{\end{eqnarray}} 
\def \to   {\mbox{$\rightarrow$}} 
\def\Re{\mathop{\rm Re}}
\def\ord#1{{\cal O}\(#1\)}
\def\epem{e^+e^-}
\def\ep{\epsilon}
\def\e{\epsilon}
\def\asb{{}\ifmmode \bar{\alpha}_s \else $\bar{\alpha}_s$\fi}
\def\Dff{{\cal D}}
\def\Dnp{{\cal D}_{\scriptscriptstyle \rm NP}^B}

\newcommand\eq[1]{Eq.~(\ref{#1})}

\def\lq{\left[} 
\def\rq{\right]} 
\def\rg{\right\}} 
\def\lg{\left\{} 
\def\({\left(} 
\def\){\right)} 

\def\mpole{m}

  \newcommand{\ccaption}[2]{
    \begin{center}
    \parbox{0.85\textwidth}{
      \caption[#1]{\small{{#2}}}
      }
    \end{center}
    }
\catcode`\@=11

\@addtoreset{equation}{section}
\def\theequation{\thesection.\arabic{equation}}

\def\section{\@startsection{section}{1}{\z@}{3.5ex plus 1ex minus
   .2ex}{2.3ex plus .2ex}{\large\bf}}

\def\thesection{\arabic{section}}

\def\appendix{\setcounter{section}{0}
 \def\thesection{\Alph{section}}
 \def\theequation{\Alph{section}.\arabic{equation}}
 \def\section{\@startsection{section}{1}{\z@}{3.5ex plus 1ex minus
   .2ex}{2.3ex plus .2ex}{\large\bf}}
 \def\subsection{\@startsection{section}{2}{\z@}{3.25ex plus 1ex minus
   .2ex}{1.5ex plus .2ex}{\large\bf}}}

\relax

\catcode`\@=12

\begin{document}
\begin{titlepage}
\nopagebreak
{\flushright{
        \begin{minipage}{4cm}
        Bicocca-FT-03-7\hfill\\
	DCPT/03/32\hfill\\
        DESY 03-027\hfill\\
        IPPP/03/16\hfill\\
  \end{minipage}        }

}
\vfill
\begin{center}
{\LARGE 
 \bf \sc
On a possible measurement of $\as$ from $B$-$\overline{B}$ correlations
in $Z^0$ decay

}
\vskip .5cm

{\large A.~Brandenburg$^1$, P.~Nason$^2$ and C.~Oleari$^3$}

\vskip .5cm
{\sl
$^1$DESY-Theorie, 22603 Hamburg, Germany\\
$^2$INFN, Sezione di Milano, Piazza della Scienza 3,
20126 Milan, Italy\\
$^3$Department of Physics, IPPP, South Road, Durham DH1 3LE, UK
}

\end{center}
\nopagebreak
\vfill
\begin{abstract}

Motivated by recent preliminary
results from the SLD Collaboration on the measurement of
angle-dependent $B$-$\overline{B}$ energy correlations in $Z^0 \,\to\, b
\bar{b}$ events, we propose a class of observables that
can be computed as a power
expansion in the strong coupling constant $\as$, of order $\as$ at the Born
level and that can 
be used for a precision measurement of $\as(M_Z)$. We
compute their next-to-leading order $\ord{\as^2}$ corrections in the strong
coupling constant, including exactly quark-mass effects.
We show that, in the theoretical evaluation of these quantities,
large logarithms of
the ratio of the mass of the final quark over the centre-of-mass energy
cancel out. Thus, these variables have a well-behaved perturbative expansion
in $\as(M_Z)$.  We study the theoretical
uncertainties  due to the renormalization-scale dependence and the quark-mass
scheme and we address the question of which mass scheme is more appropriate
for these variables. 
\end{abstract}
\vskip 1cm
\vfill
\end{titlepage}

\section{Introduction}
\label{sec:intro}

Hadronic final states in $e^+e^-$ annihilation have been intensively studied
in recent years, both at LEP and SLC, in order to perform QCD tests and to
determine $\as$~\cite{Hinchliffe:2002ex}.
Interesting studies on bottom flavoured hadronic events have also
been performed, in order to test the flavour independence of the strong
coupling constant and to  search for ``running mass''
effects~\cite{Barate:2000ab,Abreu:1998ey,Abbiendi:1999fs,Abbiendi:2001tw,Abe:1998kr,Brandenburg:1999nb}. 
However, up to now, heavy flavoured jets have been studied with the same
tools used to study light flavoured jets.
For example, the same shape
variables used for light-quark jets (like thrust, cluster multiplicities, etc.)
have been applied to heavy flavoured events.
On the other hand, heavy flavoured events may have some advantages,
because the kinematic distributions
of the produced heavy quark can be computed theoretically with much better
accuracy than the corresponding light-quark quantities. Furthermore,
the kinematic of the final-state heavy flavoured hadron is more
closely related to that of the final-state quark, since hadronization effects
involve energy scales that are much smaller than the quark mass.

Recently, the SLD Collaboration~\cite{Abe:2002iw} has published
preliminary data on the double inclusive
$B$-$\overline{B}$ production, and studied quantities (proposed in
Ref.~\cite{Burrows:1992uh}) that carry
information about the angular dependence of the $B$-$\overline{B}$
energy correlations.

The main purpose of the present work is to demonstrate that
a slight variant of the quantities introduced
in Ref.~\cite{Burrows:1992uh} has suitable properties to perform
a novel, heavy-flavour-specific determination of the strong coupling
constant. We will demonstrate the following facts:
\begin{itemize}
\item The observables we are considering have a well-defined, finite expansion in
the strong coupling constant.
\item Non-perturbative corrections of order $\LambdaQCD/m$,
where $m$ is the heavy-quark mass, cancel
in these variables. The only remaining non-perturbative
corrections are of order $\LambdaQCD/\ecm$,
where $\ecm$ is the centre-of-mass energy.
\item Potentially large terms, enhanced by powers of $\log (\ecm/m)$,
cancel at all orders in perturbation theory,
provided $\as$ is evaluated at
a scale near $\ecm$. This cancellation holds as long as one can
neglect multiple heavy-flavour pair production in our process,
which is certainly the case for $b$ production in $Z^0$ decays.
\end{itemize}
The quantities we are considering have a well defined massless
limit, provided 
one can neglect multiple heavy-flavour pair production.
This limitation is what makes these quantities heavy-flavour specific.
In fact, only for heavy flavours multiple pair production is truly
negligible. Thus, although these quantities have formally a well defined
massless limit, they cannot be used in a straightforward way to study
light-flavour production.

%
Tools to compute heavy-flavour production at the next-to-leading-order (NLO)
level have been available in the literature for the past
few  years~\cite{Rodrigo:1997gy,Rodrigo:1999qg,Bernreuther:1997jn,
Brandenburg:1998pu,Nason:1998nw}.
We have used our
calculations~\cite{Bernreuther:1997jn,Brandenburg:1998pu,Nason:1998nw} 
in order to compute the NLO corrections to the observables we
propose.

In Sec.~\ref{sec:double_inclusive}, we define the variables we are
considering and
show that their perturbative expansion in terms of the strong coupling
constant $\as\(M_Z\)$ is well behaved, and free from large logarithms of the
ratio of the quark mass over the centre-of-mass energy.  The SLD
Collaboration has provided us with (preliminary) data for these new
variables~\cite{newSLD}, and thus we have been able to check our
NLO predictions against the data.

In Sec.~\ref{sec:m_to_zero} we investigate the massless limit of these
variables, and 
show explicitly that potentially large logarithms cancel out.

In Sec.~\ref{sec:th_uncertainties} we study theoretical uncertainties of our
results. We analyze the dependence of the NLO results on the renormalization
scale and compare the results obtained with the pole-mass definition to those
obtained using the \MSB\ mass, addressing the question of which of the two
mass definitions is more suitable for this calculation.  Finally, in
Sec.~\ref{sec:conclusions} we give our conclusions.

In Appendix~\ref{app:coefficients}, we collect some leading order (LO) formulae
used in our analysis, and in Appendix~\ref{app:as_renorm} we comment on the
renormalization scheme used in this paper.

\section{Double-inclusive heavy-quark cross section}
\label{sec:double_inclusive}
Heavy flavoured hadron production
proceeds through the process
\begin{displaymath}
e^+e^- \, \to \,
Z^0/\gamma\,\to\, Q + \overline{Q} + X\,,
\end{displaymath}
(where $Q$ is a heavy quark of mass
$m$) followed by the hadronization of the final heavy quarks.
In the following we will focus upon $B$ meson production.

Defining the $B$ hadron scaled energy $\xb=2E_{\scriptscriptstyle B}/\ecm$,
we consider the following quantities~\footnote{
The quantities $D_i$ and $D_{ij}$ were also
considered in~\cite{Burrows:1992uh} and the ratios $G_{ij}$ in~\cite{burrows}.}
\begin{eqnarray}
\label{eq:D_i_def}
D_i&=&\frac{1}{\sigB}\int \xb^{i-1} \frac{d\sigB}{d\xb}\,d\xb\,, \\ 
\label{eq:D_ij_def}
D_{ij}(\phi)&=&\frac{1}{\sigB}\int \xbo^{i-1} \xbt^{j-1}
\frac{d^3\sigB}{d\xbo d\xbt d\cos\phi} \,d\xbo d\xbt \,,
  \\
\label{eq:G_ij_def}
G_{ij}(\phi)&=&\frac{D_{ij}(\phi)}{D_i D_j} \,,
\end{eqnarray}
where $d\sigB/d\xb$ is the differential cross section for the final-state
hadron production, $\phi$ is the angle between the two hadrons, $\sigB$
the total cross section and $x_{\scriptscriptstyle B_{1/2}}$ are
the scaled energies of the two final $B$ hadrons.

The quantities studied by the authors of~\cite{Burrows:1992uh}
are related to the ratios $G_{ij}/G_{11}$ by a calculable perturbative factor
($P_iP_j/P_1^2$ in the notation of~\cite{Burrows:1992uh}).
In Ref.~\cite{Abe:2002iw} these ratios were compared to
experimental results. However, these quantities have a few drawbacks.
First of all, they depend upon a ``theoretical'' factor. It would be
preferable to deal with quantities that have an experimental definition,
free of theoretical assumptions. Second and most important,
they have a perturbative expansion that starts at leading order with
a constant. Therefore, even when corrected at the NLO level, they
would allow only for a LO determination of $\as$. Conversely,
they can be used to test the production mechanism, since in the ratios
$\as$ cancels at first approximation.

In this work, we claim that the quantities $G_{ij}$, defined
in Eq.~(\ref{eq:G_ij_def}), should instead be studied in order to
perform precision QCD tests. We will in fact show in the following
that these quantities have a perturbative expansion in $\as(\ecm)$ with
coefficients that remain finite even in the limit of large $\ecm/m$
ratios.

We write the hadronic differential cross sections as
\begin{eqnarray}
\label{eq:dsigma_B}
\frac{d\sigB}{d\xb}&=&\int  \frac{d\sigb}{d x_b}
\Dnp(x) \, \delta(\xb - x x_{b})\,d x_{b} dx\,,\\
\label{eq:dsigma3_B}
\frac{d^3\sigB}{d\xbo d\xbt d\cos\phi}
&=&\int \frac{d^3\sigb}{dx_{b_1} dx_{b_2} d\cos\phi}
\Dnp(x_1)\Dnp(x_2)  \nonumber \\
&& \mbox{} \times \delta(\xbo - x_1 x_{b_1})\,\delta(\xbt - x_2 x_{b_2})
\, dx_{b_1}  d x_{b_2} dx_1 dx_2\,,
\end{eqnarray}
where $d\sigma_b$, $x_b$, $x_{b_{1/2}}$ are the quark differential
cross section and the quark scaled energies.
The quark differential cross sections $d\sigma_b$ are calculable
order-by-order in
perturbative QCD, since the heavy-quark mass $m$ acts as a cut-off for
final-state collinear singularities.
However, in order to go from the QCD partonic cross section
$d\sigb/dx_b$
to the hadronic  one $d\sigB/d\xb$, we have 
to take into consideration non-perturbative effects, of
order $\LambdaQCD/m$, that are present in the hadronic cross section.  We
assume that all these effects are described by a non-perturbative
fragmentation function $\Dnp$.

The quark differential cross sections, although calculable,
contain logarithmically enhanced terms of the form $\as^k \log^l
\(m/\ecm\)$, so that their perturbative expansion is not well behaved
for $\ecm\gg m$. 

In the quantities $G_{ij}(\phi)$
these large logarithms cancel, together with the dependence upon
the non-perturbative fragmentation function. This is a consequence of the
factorization theorem, if one makes the additional
assumption that secondary production of 
heavy-flavour pairs is negligible, which is in fact the
case at the $Z^0$ peak.

The factorization theorem
states that mass singularities in inclusive
cross sections can be absorbed into process independent, universal
factors. In our case, mass singularities are precisely the
$\log (\ecm/m)$ terms. Thus, for $\ecm\gg m$ we have
\begin{eqnarray}
\label{eq:dsigma_b}
\frac{d\sigb}{dx_b}&=&\int  \frac{d\hat{\sigma}}{d\xhb}
\Dff_b(z) \, \delta(x_b - z \xhb)\,d \xhb dz\,,\\
\label{eq:dsigma3_b}
\frac{d^3\sigb}{dx_{b_1} dx_{b_2} d\cos\phi}
&=&\int \frac{d^3\hat{\sigma}}{d\xhbo d\xhbt d\cos\phi}
\Dff_b(z_1)\Dff_b(z_2)  \nonumber \\
&& \mbox{} \times \delta(x_{b_1} - z_1 \xhbo)\,\delta(x_{b_2} - z_2 \xhbt)
\, d\xhbo  d\xhbt dz_1 dz_2\,,
\end{eqnarray}
where the hatted cross sections do not depend upon $m$,
and all terms that are large in the $m \,\to\, 0$ limit (i.e.\ terms
proportional to powers of $\log (\ecm/m)$)
are absorbed in the fragmentation functions
$\Dff_b(z)$~\footnote{
Notice that if we had not assumed a negligible secondary heavy-flavour
pair production, we should have introduced also the gluon and light-quark
component of the heavy-quark fragmentation function.}. 
Inserting Eqs.~(\ref{eq:dsigma_b}) and~(\ref{eq:dsigma3_b})
into Eqs.~(\ref{eq:dsigma_B}) and~(\ref{eq:dsigma3_B}) and considering the
definitions in Eqs.~(\ref{eq:D_i_def})  and~(\ref{eq:D_ij_def}), we get
\beqn
D_i &=& \( \int x^{i-1} \Dnp(x)\, dx\)
\(\int z^{i-1} \Dff_b(z) \,dz \) \,\,
\frac{1}{\sigB} \int \xhb^{i-1} \frac{d\hat{\sigma}}{d\xhb}\, d\xhb,\\
D_{ij}(\phi) &=& 
\( \int x_1^{i-1} \Dnp(x_1)\, dx_1\)
\( \int x_2^{j-1} \Dnp(x_2)\, dx_2\) \nonumber\\
&& \times 
\( \int z_1^{i-1} \Dff_b(z_1) \, dz_1 \)
\( \int z_2^{j-1} \Dff_b(z_2) \, dz_2 \)  \nonumber\\
&& \times \frac{1}{\sigB} \int \xhbo^{i-1} \xhbt^{j-1}
\frac{d^3\hat{\sigma}}{d\xhbo d\xhbt d\cos\phi} \, d\xhbo  d \xhbt \,,
\eeqn
and we see that, in the ratio defining $G_{ij}(\phi)$ (see
Eq.~(\ref{eq:G_ij_def})), 
the $\Dff_b$ factors (that contain mass
singularities) cancel out, leaving a finite expression~\footnote{Of course,
this holds if we neglect terms that are suppressed 
by powers of $m$. Terms suppressed by powers of $m$ and enhanced
by powers of $\log (\ecm/m)$ may still (and in fact are) present in $G_{ij}$.
Furthermore, uncalculable effects of order $\LambdaQCD/\ecm$ may
still be present.}.
%

Since non-perturbative effects cancel in the definition of $G_{ij}$,
from now on we will make no distinction between the quantities
$D_i$, $D_{ij}$ defined at the quark and at the hadron level.
With the available codes for the calculation of NLO corrections in the
production of heavy quarks~\cite{Brandenburg:1998pu,Nason:1998nw}
at $\epem$ colliders~\footnote{A version of the program described
in~\cite{Brandenburg:1998pu} was used which implements the dipole subtraction
method for 
massive partons~\cite{Catani:2002hc} to combine real and virtual
corrections.}, we can compute the contributions up to order $\as^2$ to
the quantities $D_i$, $D_{ij}(\phi)$ and $G_{ij}(\phi)$.
We can write the perturbative expansions for $D_i$ and $D_{ij}(\phi)$ for
massive final-state quarks, renormalized in the pole-mass scheme, in the
region away from the two-jet final state (the two parton final state gives a
contribution proportional to $\delta(\phi-\pi)$) as
\beqn
D_i &=& 1-a_i\frac{\as(\mu)}{2\pi} + \ord{\as^2}\,,\\
\label{eq:D_ij_expans}
D_{ij}(\phi)&=&B_{ij}(\phi)\frac{\as(\mu)}{2\pi}+
\lq C_{ij}(\phi)+2\pi\, b_0\, B_{ij}(\phi)\log\(\frac{\mu^2}{\ecm^2}\) \rq
\left(\frac{\as(\mu)}{2\pi}\right)^2 \nonumber\\ && \mbox{}+ {\cal O}(\as^3),
\qquad \qquad {\rm for\ } \phi \neq \pi\,,
\eeqn
so that the expression for $G_{ij}$ becomes
\beqn
\label{eq:G_ij_expans}
G_{ij}(\phi) &=& B_{ij}(\phi)\frac{\as(\mu)}{2\pi}+
\Bigg\{ C_{ij}(\phi)  \nonumber\\ && \mbox{}+ 
\lq2\pi \, b_0\log\(\frac{\mu^2}{\ecm^2}\)+a_i+a_j\rq B_{ij}(\phi)\Bigg\}
\left(\frac{\as(\mu)}{2\pi}\right)^2 
+ {\cal O}(\as^3)\,,
\eeqn
where
\beq
 b_0 = \frac{11 C_A - 4 n_f T_F}{12\pi},
\eeq
$\mu$ is the renormalization scale, $C_A = N_c = 3$ for an SU(3) gauge
theory, $T_F=1/2$ and $n_f$ is the number of active flavours. 

We observe that expanding the ratio in powers of $\as$ is essential
to implement the cancellation of large logarithms in $G_{ij}$
order-by-order in the perturbative series. If we had not
done so, the numerator and denominator of $G_{ij}$ would contain large
logarithms at some fixed order, and their ratio would be misleading 
(i.e.\ it would be sensitive to higher order terms with higher powers
of logarithms).

\begin{figure}
\begin{center}
\epsfig{file=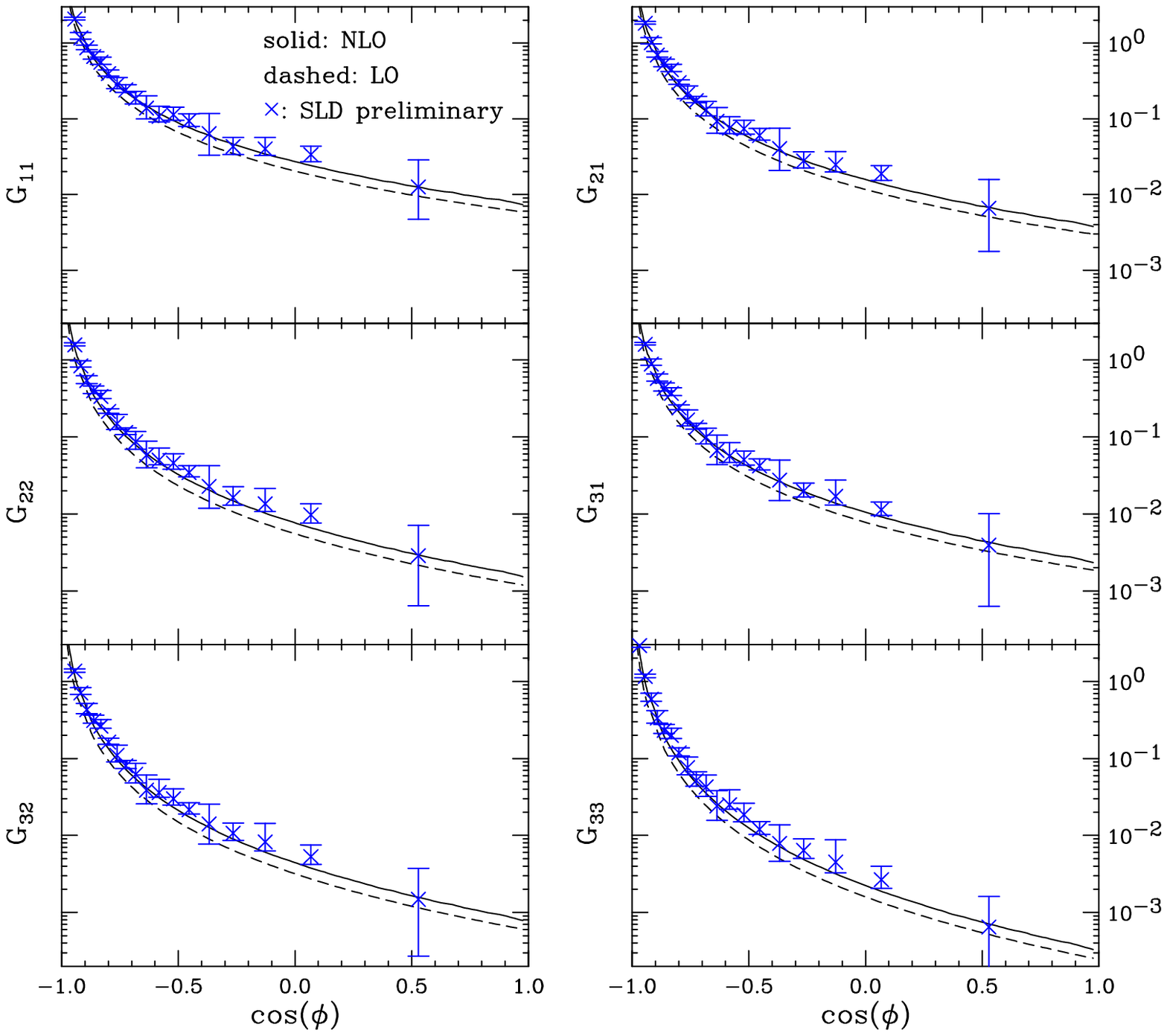,width=1\textwidth,clip=}
\ccaption{}
{\label{fig:G_ij} 
Leading order (dashed) and next-to-leading order (solid) curves for
$G_{ij}$, $i,j=1,2,3$, at a renormalization scale $\mu=\ecm=91.2$~GeV, using
$\as(\ecm)=0.12$.
The error bars in the preliminary data from the SLD
Collaboration~\cite{newSLD} are the 
quadratic sum of systematic and statistical errors. } 
\end{center}                                  
\end{figure}
In Fig.~\ref{fig:G_ij} we have plotted $G_{ij}$ at leading  and
next-to-leading order in $\as$, for $i,j=1,2,3$, at a renormalization
scale $\mu=\ecm=91.2$~GeV, using $\as(\ecm)=0.12$, with $n_f=5$ active
flavours.  The 
NLO curves are in good agreement with the SLD preliminary data.

Conversely, the SLD data can be used to extract the value of $\as$. In fact,
from the knowledge of the numerical values of the coefficients of the $\as$
and $\as^2$ contributions in \eq{eq:G_ij_expans}, we can fit $G_{ij}(\phi)$
to the data, keeping $\as$ as a free parameter.

We have computed tables for the $\as$ and $\as^2$
coefficients of Eq.~(\ref{eq:G_ij_expans}),
 that can be used in a more detailed analysis. These tables
(and the programs to generate them) can be obtained from the authors.

\subsection{Secondary $\boldsymbol{b}$-quark and the running coupling}
We should mention here that, in the calculations of
Ref.~\cite{Nason:1998nw}, the $b$ quark is treated as a
heavy particle, and thus does not contribute to the running of the coupling
constant. Thus, in principle, one should use this calculation in
conjunction with the 4-flavour coupling constant $\as^{(4)}$. Since we are
dealing with $Z^0$ decays, however, secondary $b$-quark pairs may in fact be
produced (from gluon splitting), and when this happens, they behave, in some
sense, as light flavours, contributing terms that effectively modify
the running of $\as$.

\begin{figure}[htb]
\begin{center}
\epsfig{file=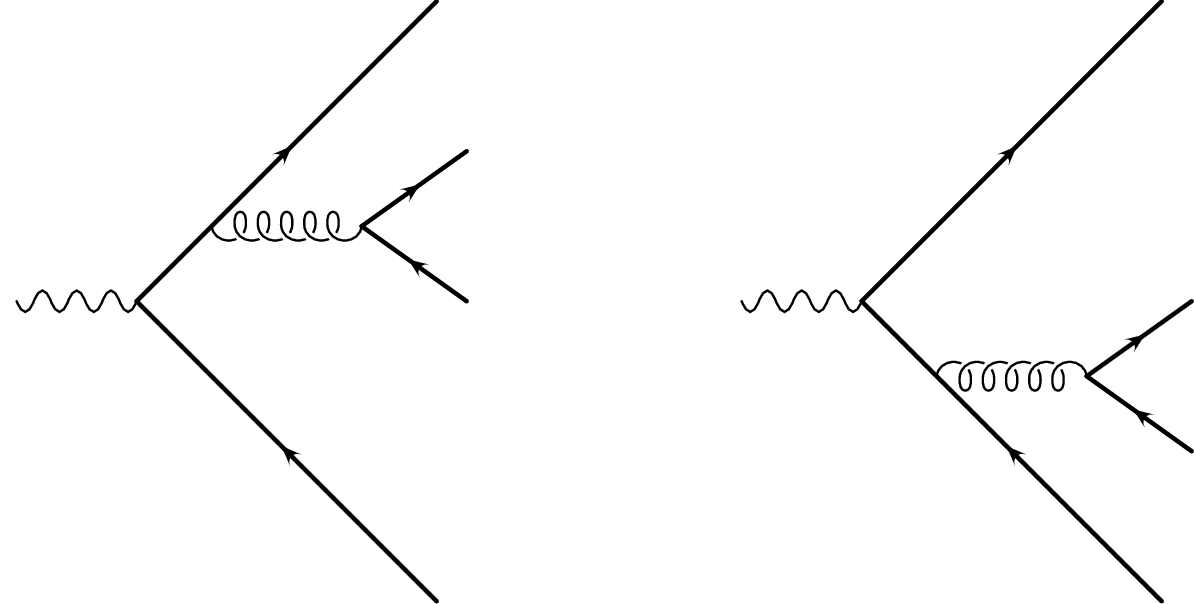,width=0.5\textwidth}
\ccaption{}
{\label{fig:secondaryb}
Secondary $b$-quark production.}
\end{center}
\end{figure}

In Fig.~\ref{fig:secondaryb}, we show the Feynman graphs for secondary
heavy-flavour production, giving rise to four massive-quark final state.
The graphs in Fig.~\ref{fig:secondaryb} also appear with the two
identical quarks and/or the two identical antiquarks exchanged.  When taking
the amplitude squared, this gives rise to two interference terms. Their
contribution is known to be extremely small, and will be neglected in the
following.

In addition, virtual $b$-quark loops arise only in the self-energy insertion
of the emitted (on-shell) gluon.  Their contribution is zero in the
renormalization scheme we use, since, in this scheme, the heavy flavour
decouples for small momenta (see Appendix~\ref{app:as_renorm} for more
details).  Thus, the only effect of the heavy flavour is given by the graphs
of Fig.~\ref{fig:secondaryb}.

In the quantities we are considering, the possibility of detecting secondary
(instead of primary) $B$ mesons has little impact, since secondary $B$'s are
very soft.  If we make the approximation of neglecting effects that are
suppressed by powers of the mass in
the secondary $b$ line, the contribution of the graphs of
Fig.~\ref{fig:secondaryb} (without interference) is easily included using the
following prescription:
\begin{itemize}
\item[-] compute the cross section with the number of light flavours $n_{lf}$
  equal to 5;
\item[-] use the 5 flavour coupling $\as^{(5)}$;
\item[-] do not include any other graph with secondary heavy flavours.
\end{itemize}
In essence, the prescription amounts to including the secondary heavy flavour
by simply increasing $n_{lf}$ by one. When doing this, one in fact is also
changing the renormalization scheme for the heavy flavour, by treating it as
if it were light.  Thus, the correct coupling constant to use in this
framework is $\as^{(5)}$.
In Appendix~\ref{app:as_renorm} we discuss this change
of scheme in more detail and from a slightly different point of view.

Secondary $b$-quark production may also arise from primary light quarks,
connected with the $Z^0/\gamma$ vertex. We
neglect altogether this mechanism, since it generates only soft $b$ quarks.

\section{The $\boldsymbol{m\,\to\,0}$ limit}
\label{sec:m_to_zero}
As explained in Sec.~\ref{sec:double_inclusive}, $G_{ij}(\phi)$ are free from
potentially dangerous logarithms of the ratio $\ecm/m$.  In order to
investigate the stability of our expressions in the $m\,\to\,0$ limit, we
introduce the integral of $G_{ij}$ over $\phi$ and we define (we drop the
$B/b$ subscript for ease of notation)
\begin{eqnarray}
\label{eq_r_ij_def}
r_{ij} = \int_{-1}^{1} G_{ij}(\phi)\,d\cos\phi =
{\langle x_1^{i-1} x_2^{j-1}\rangle\over 
\langle x_1^{i-1}\rangle\langle x_2^{j-1}\rangle}\,,
\end{eqnarray}
where
\beqn
\langle x_1^{i-1}\rangle &=& \langle x_2^{i-1}\rangle \equiv D_i\,,\\
\langle x_1^{i-1} x_2^{j-1}\rangle &\equiv& \int_{-1}^1 D_{ij}(\phi) 
\,d\cos\phi\,.
\eeqn
To evaluate Eq.~(\ref{eq_r_ij_def}) up to order $\as^2$ and for 
$i,j=2,3$,  we rewrite $r_{ij}$ as follows
\begin{eqnarray}
r_{ij} &=& 1+{\langle x_1^{i-1} x_2^{j-1}\rangle-
\langle x_1^{i-1}\rangle\langle x_2^{j-1}\rangle\over 
\langle x_1^{i-1}\rangle\langle x_2^{j-1}\rangle}\nonumber \\ 
\label{eq_r_ij_manip}
&=& 
1+{\langle (1-x_1^{i-1})(1- x_2^{j-1})\rangle-
\langle 1-x_1^{i-1}\rangle\langle 1- x_2^{j-1}\rangle\over 
\langle x_1^{i-1}\rangle \langle  x_2^{j-1}\rangle }\,,
\end{eqnarray}
and we expand in $\as$
\begin{eqnarray}
\label{eq:a_i}
\langle 1-x_1^{i-1}\rangle&=&
\langle 1- x_2^{i-1}\rangle =  a_i\,
{\as(\mu) \over 2\pi} +{\cal O}
\left(\as^2\right),\\
\label{eq:b_ij}
\langle (1-x_1^{i-1})(1- x_2^{j-1})\rangle&=& b_{ij}\, {\as(\mu) \over 2\pi}+
\lq c_{ij} + 2\pi \,b_0\, b_{ij} \log\(\frac{\mu^2}{\ecm^2}\)  \rq
\left({\as(\mu) \over 2\pi}\right)^2+ \ord{\as^3},\nonumber\\
\end{eqnarray}
to obtain
\beq
\label{eq:r_ij_final}
r_{ij} = 1 + b_{ij}\,{\as(\mu) \over 2\pi} 
+\left[c_{ij}+
\(a_i+a_j\)b_{ij}-a_ia_j
+ 2\pi\, b_0\,b_{ij} \log\(\frac{\mu^2}{\ecm^2} \)
\right] \left({\as(\mu) \over 2\pi}\right)^2
+\ord{\as^3} .
\eeq
Note that, in Eq.~(\ref{eq_r_ij_manip}), the expectation value of
$x_{1}^{i-1}$ and $x_{2}^{j-1}$ 
should be known only up to order $\as$, and that, in the correlation term
$\langle (1-x_1^{i-1})(1- x_2^{j-1})\rangle$, 
the two-jet final state gives zero contribution, since
it is  proportional to $\delta(1-x_1)\delta(1- x_2)$, at all orders in $\as$.
This is the reason why we do not need to know the $\as^2$ two-loop virtual
term for $\epem\,\to\,Q \overline{Q}$.

The coefficients $a_i$ and $b_{ij}$ can be computed analytically. We have
collected their expressions in Appendix~\ref{app:coefficients}.
\begin{table}[ht]
\caption{Results for the coefficients $a_i$, $b_{ij}$ and $c_{ij}$ appearing
  in the definition of $r_{ij}$ (see
  Eqs.~(\ref{eq:a_i})--(\ref{eq:r_ij_final})). The mass $m$ is in GeV and
  $\mu=\ecm=91.2$~GeV. } 
\begin{center}
\begin{tabular}{
|c|c|c|c|c|c|c|}\hline
& $m=0.01$  & $m=0.1$
& $m=1$  & $m=3$   & $m=5$   & $m=10$   \\  \hline
$a_2$ 
& 29.1611  & 20.9741 
& 12.7866  & 8.87675 & 7.05412  & 4.56726 \\ 
$a_3$ 
& 46.6475 & 33.8553 
& 21.0574  & 14.9209 & 12.0317 & 8.01007  \\   
$b_{22}$ 
& 0.666666 & 0.666589  
& 0.661895 & 0.636326 & 0.598452&  0.478637\\
$b_{32}$
& 1.13333 & 1.13321 
& 1.12569 & 1.08448 & 1.02307  & 0.826778\\
$b_{33}$ 
& 1.91111 & 1.91093
& 1.89945 & 1.83563 & 1.73915 &  1.42415\\
$c_{22}$ 
& 820.1(3) & 420.51(4) 
& 155.14(1)  & 75.787(7) & 49.176(3) & 22.943(1) \\
$c_{32}$ 
&1290.5(4) & 663.99(7)   
& 247.19(2)  & 122.15(1) & 80.053(6) & 38.268(3) \\
$c_{33}$ 
&2027.5(5) & 1046.5(1) 
& 393.09(3) & 196.60(2) & 130.23(1)  & 63.871(5)\\
\hline
\end{tabular}
\label{tab:nlo_coeff}
\end{center}
\end{table}    
In Table~\ref{tab:nlo_coeff}, we give the numerical values for the
coefficients $a_i$, $b_{ij}$ and $c_{ij}$, for a down-type quark with mass
$m = 0.01$, $0.1$, $1$, $3$, $5$ and $10$~GeV, at a center-of-mass energy 
$\ecm = \mu = 91.2$~GeV. 
\begin{table}[ht]
\caption{Results for $r_{ij}$. The mass $m$ is in GeV and
  $\mu=\ecm=91.2$~GeV.} 
\begin{center}
\begin{tabular}{
|c|c|c|c|}\hline
$m$
& 
$r_{22}-1$ & $r_{32}-1$   & $r_{33}-1$    \\  \hline
0.01 & 0.106103\*\,$\as$+0.215(4)\*\,$\as^2$
     & 0.180375\*\,$\as$+0.403(6)\*\,$\as^2$
     & 0.304162\*\,$\as$+0.750(9)\*\,$\as^2$\\
0.1 & 0.106091\*\,$\as$+0.217(1)\*\,$\as^2$
    & 0.180356\*\,$\as$+0.407(2)\*\,$\as^2$ 
    & 0.304133\*\,$\as$+0.755(3)\*\,$\as^2$ \\ 
1 & 0.105344\*\,$\as$+0.2175(4)\*\,$\as^2$
  & 0.179159\*\,$\as$+0.4066(7)\*\,$\as^2$
  & 0.302307\*\,$\as$+0.7524(1)\*\,$\as^2$ \\   
3 & 0.101274\*\,$\as$+0.2100(1)\*\,$\as^2$
  & 0.172600\*\,$\as$+0.3930(3)\*\,$\as^2$
  & 0.292149\*\,$\as$+0.7287(5)\*\,$\as^2$\\
5 & 0.0952466\*\,$\as$+0.19905(9)\*\,$\as^2$
  & 0.162827\*\,$\as$+0.3725(2)\*\,$\as^2$
  & 0.276794\*\,$\as$+0.6919(3)\*\,$\as^2$ \\
10 & 0.0761774\*\,$\as$+0.16351(4)\*\,$\as^2$
   & 0.131586\*\,$\as$+0.30605(7)\*\,$\as^2$ 
   & 0.226661\*\,$\as$+0.5706(1)\*\,$\as^2$\\
\hline
\end{tabular}
\vspace*{1em}
\label{tab:r_ij}
\end{center}
\end{table} 
Observe that, while the individual coefficients $a_i$ and $c_{ij}$ can be
quite large, due to the presence of logarithmic terms, the ratio $r_{ij}$ is
well behaved, as illustrated
in Table~\ref{tab:r_ij}. This is due to the fact that the large logarithms
present in the single coefficients cancel out in the ratio $r_{ij}$.

\section{Theoretical uncertainties} 
\label{sec:th_uncertainties}
In this section, we investigate the theoretical uncertainties related to the
renormalization scale and mass scheme used in the calculation.

The renormalization-scale dependence of $r_{ij}$
is illustrated in Table~\ref{tab:renorm} for the case $m=5$~GeV, using
the two-loop evolution of $\as$ with  $\as=0.12$ for 
$\mu=\ecm=91.2$~GeV. For $\ecm/2<\mu<2\ecm$ the variation of $(r_{ij}-1)$
around the central value  $\mu=\ecm$ is roughly $\pm 10\%$ at leading order.
It is reduced to $\pm 5\%$ at NLO.

\begin{table}[ht]
\caption{Results for $r_{ij}$ for different renormalization scales with 
 $\as(\ecm)=0.12$, $\ecm=91.2$~GeV and $m=5$~GeV. }
\begin{center}
\begin{tabular}{
|c|c|c|c|c|c|c|}\hline
& \multicolumn{2}{|c|} {$\mu=\ecm/2$} &  \multicolumn{2}{|c|} {$\mu=\ecm$} &   
\multicolumn{2}{|c|} {$\mu=2\ecm$} \\
& LO & NLO & LO & NLO & LO & NLO \\
\hline
$r_{22}-1$ & 0.0127974 & 0.014937(2) & 0.0114296 & 0.014296(1) 
           & 0.0103320 & 0.013622(1)\\
$r_{32}-1$ & 0.0218776 & 0.026116(3)& 0.0195393 & 0.024903(2) 
           & 0.0176629 & 0.023667(2)\\
$r_{33}-1$ & 0.0371903 & 0.045455(5)& 0.0332153 & 0.043179(4) 
           & 0.0300257 & 0.040922(3)\\
\hline
\end{tabular}
\vspace*{1em}
\label{tab:renorm}
\end{center}
\end{table}  

In order to investigate the mass-scheme dependence, 
we now explicitly distinguish between the pole mass $\mpole$
and the \MSB\ mass $\overline{m}(\mu)$ (also called running mass).
Using the relation between $\mpole$ and $\overline{m}(\mu)$
\beq
\mpole = \overline{m}(\mu) \lg 1 + \frac{\as(\mu)}{2\pi} C_F 
\lq2-\frac{3}{2} \log\(\frac{\overline{m}(\mu)^2}{\mu^2}\)\rq + \ord{\as^2}
\rg\,, 
\eeq
and defining 
\beq
z = \(\frac{\mpole}{\ecm}\)^2\,,  \qquad \qquad
\bar{z}(\mu) = \(\frac{\overline{m}(\mu)}{\ecm}\)^2\,,
\eeq
we have 
\beq
\label{eq:rijmsbar}
\bar{r}_{ij}\(\bar{z}\) \equiv
r_{ij}\(\bar{z}\(\mu\)+\Delta\bar{z}\(\mu\)\)
=r_{ij}(\bar{z}(\mu))+\left({\as(\mu)\over 2\*\pi}\right)^2
\*\Delta b_{ij}\(\bar{z}\(\mu\)\)+ \ord{\as^3}\,,
\eeq
with
\beq
\label{msbarmass}
\Delta b_{ij}(\bar{z}(\mu))=4\*C_F\*\bar{z}(\mu)\*
\left[1-{3\over 4}\ln\left(\bar{z}(\mu){\*\ecm^2 \over \mu^2}\right)
\right]\*{db_{ij}\over dz}(\bar{z}(\mu))\,.
\eeq

We have evaluated  Eq.~(\ref{eq:rijmsbar}) for
$\overline{m}(\ecm)=3$~GeV using the analytic expressions for the
coefficients $b_{ij}$ collected in Appendix~\ref{app:coefficients}.
The value $\overline{m}(\ecm)=3$~GeV corresponds
roughly to what one gets by converting $\mpole$ at the $b$-mass scale to
$\overline{m}$, and then evolving $\overline{m}$ up to $\mu=\ecm$.  We have
further studied the dependence of 
Eq.~(\ref{eq:rijmsbar}) on the renormalization scale $\mu$. We
have used the 2-loop evolution of $\overline{m}(\mu)$, which gives
$\overline{m}(\ecm/2)=3.20$~GeV and $\overline{m}(2\ecm)=2.83$~GeV. The
results for $(\bar{r}_{ij}-1)$ are listed in
Table~\ref{tab:renorm_msbar}. Using 
the \MSB\ definition for the mass, the theoretical uncertainties due to the
scale ambiguity are practically of the same size as in the pole-mass scheme.

\begin{table}[ht]
\caption{Results for $\bar{r}_{ij}$ for different renormalization scales with
 $\as(\ecm)=0.12$, $\ecm=91.2$~GeV and
$\overline{m}(\ecm)=3$~GeV. }
\begin{center}
\begin{tabular}{
|c|c|c|c|c|c|c|}\hline
& \multicolumn{2}{|c|} {$\mu=\ecm/2$} &  \multicolumn{2}{|c|} {$\mu=\ecm$}
&
\multicolumn{2}{|c|} {$\mu=2\ecm$} \\
& LO & NLO & LO & NLO & LO & NLO \\
\hline
$\bar{r}_{22}-1$ & 0.0135360 & 0.015444(2) & 0.0121529 & 0.014884(2)
           & 0.0110323 & 0.014265(2) \\
$\bar{r}_{32}-1$ & 0.0230755 & 0.026986(4)& 0.0207120 & 0.025897(4)
           & 0.0187981 & 0.024743(3)\\
$\bar{r}_{33}-1$ & 0.0390735 & 0.046904(8)& 0.0350579 & 0.044811(7)
           & 0.0318083 & 0.042675(5)\\
\hline
\end{tabular}
\vspace*{1em}
\label{tab:renorm_msbar}
\end{center}
\end{table}

Comparing the results for $r_{ij}$ obtained with an on-shell mass
$\mpole=5$~GeV with those obtained with an \MSB\ mass
$\overline{m}(\ecm) = 3$~GeV, we found that, at leading order, the
differences between $(r_{ij}-1)$ using the two different mass definitions
amount to $5\div 6$\%.  This theoretical uncertainty is reduced to less than
4\% at NLO.

\begin{figure}
\begin{center}
\epsfig{file=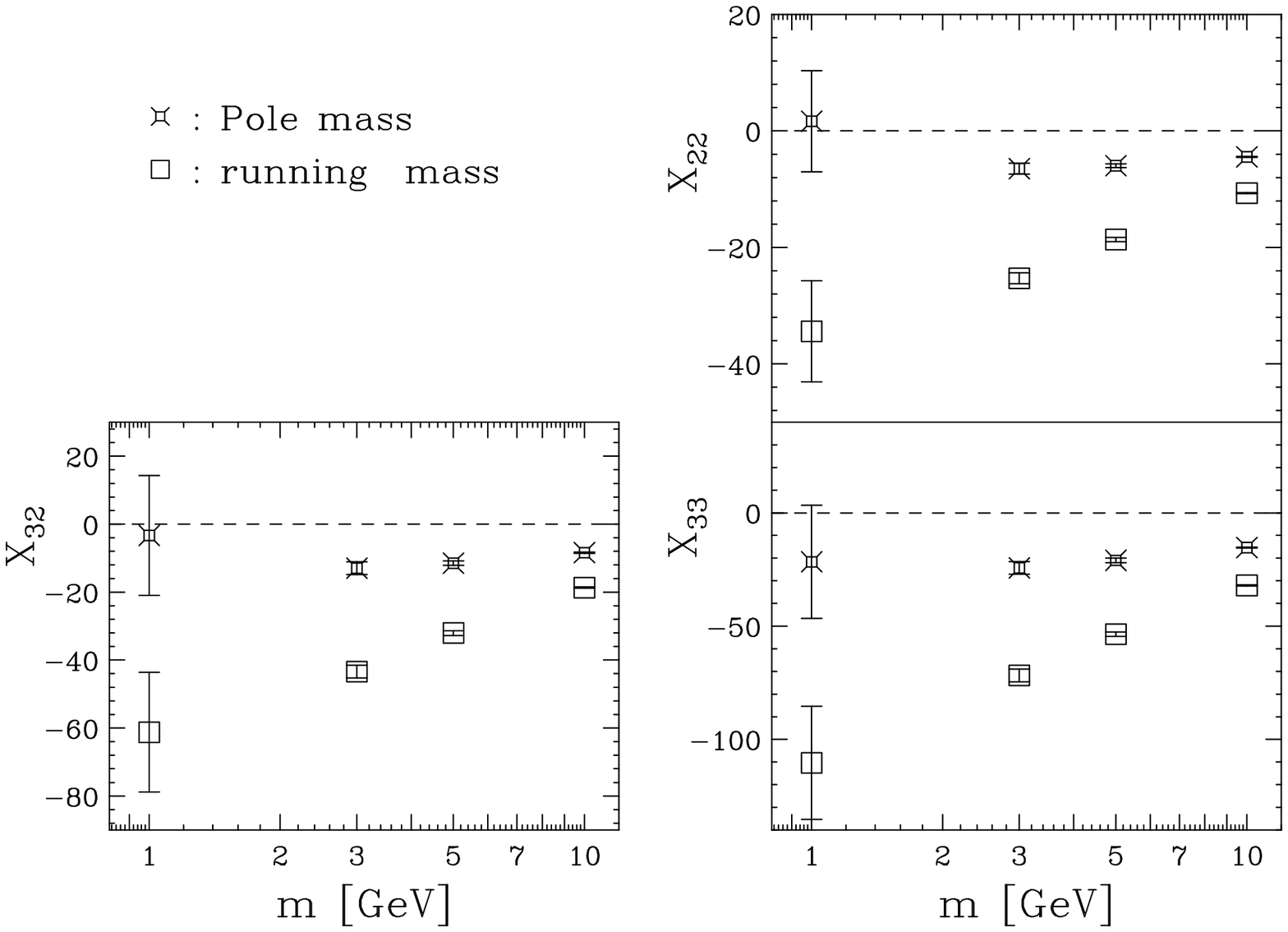,width=0.9\textwidth,clip=}
\caption{\label{fig:runningmass} 
The quantities $X_{ij}$, defined in Eq.~(\ref{eq:def_X_ij}), plotted as a
function of $m$, 
both in the pole-mass scheme and in the \MSB-mass scheme, with
$\ecm=\mu=91.2$~GeV.}
\end{center}                                  
\end{figure}

An interesting question can be raised about the use
of the pole mass versus the \MSB\ mass,
that is to say, which
one is more natural in this context. In practice, this is
a very difficult question to answer. In inclusive processes, in general,
the \MSB\ mass is better, since it accounts for large powers
of logarithms of $m/\ecm$ multiplying powers
of the mass at all orders in perturbation theory.
The case we are considering, however, is not an inclusive
process, since we are weighting the cross section with final-state kinematic
variables. We are thus unable to give 
a full answer to this question. However, we can
certainly ask whether, at the order we are considering,
there is some numerical indication that using the \MSB\ mass
accounts at least for a large part of the logarithmic terms
in our expression.
The mass dependence of the coefficients of $r_{ij}$ is reported 
in Table~\ref{tab:r_ij}.
Calling $r_{ij}^{(2)}$ the coefficient of $\as^2$ in $r_{ij}$, we
consider the expression
\begin{equation}
\label{eq:def_X_ij}
X_{ij}(z)= \frac{r_{ij}^{(2)}(z)-r_{ij}^{(2)}(0)}{z}
\end{equation}
as a function of $\log z$.
The same expression in the \MSB\ scheme is given by
\begin{equation}
\overline{X}_{ij}(\bar{z})=
\frac{r_{ij}^{(2)}(\bar{z})-r_{ij}^{(2)}(0)}{\bar{z}}
+ \frac{4}{(2\pi)^2}
C_F\left[1-\frac{3}{4}\log\bar{z} \right]\frac{d b_{ij}}{dz}(\bar{z})\;,
\end{equation}
where we use $\mu=\ecm$, and $\bar{z}=\bar{z}(\mu)$. If the slope
of the second expression versus $\log m$ is smaller
than the slope of the first one, we can take this as an indication of the
fact that it is more appropriate to use the \MSB\ mass instead of the pole
mass. The results are displayed in Fig.~\ref{fig:runningmass}.
The value $r_{ij}^{(2)}(0)$ has been taken to correspond
to $m=0.1$~GeV in Table~\ref{tab:r_ij}
(the point $m=0.01$~GeV has been excluded, because of the large errors).
The points at $m=1$, 3, 5 and 10~GeV are then plotted. 
It is quite clear from the figure that the logarithmic slope
is smaller for the pole-mass scheme. Furthermore, the radiative correction
to the mass term is always smaller for the pole-mass scheme than for the
\MSB\ scheme. We thus conclude that the quantities $r_{ij}$ seem to be
more naturally described in the pole-mass scheme rather than
in the \MSB-mass scheme.

\section{Conclusions}
\label{sec:conclusions}
In this paper we have investigated new measurable quantities
(related to the quantities introduced in Ref.~\cite{Burrows:1992uh})
defined in $\epem$
annihilation, that carry information about the angle-dependent
$B$-$\overline{B}$ energy correlation.
We have shown that these variables have
 a well-behaved expansion in terms of the
strong coupling constant $\as$, since large logarithms in the ratio of the
mass of the final quarks over the  centre-of-mass-energy cancel out order
by order, together with non-perturbative effects of order $\LambdaQCD/m$.

We have compared the computed NLO results with the available preliminary
data from the SLD Collaboration~\cite{newSLD}, and found good agreement.  
In addition, we have generated
tables~\footnote{The programs and the tables can be obtained upon request
from the authors.}
that can be used in a fit to the data,
in order to extract the strong coupling constant $\as$ at
next-to-leading order.

We have investigated the renormalization-scale and mass-scheme
dependence of our results, and
found that, varying the renormalization scale $\mu$ in the range
$\ecm/2<\mu<2\ecm$, the scale dependence of $(r_{ij}-1)$ is reduced from
roughly $\pm 10\%$ at leading order to $\pm 5\%$ at next-to-leading order.
The theoretical uncertainties related to the mass definition are less than
4\% at NLO.
We have shown that these quantities seem to be more naturally described
in the pole-mass rather than in the \MSB-mass scheme.

From the discussion given in this paper, one may wonder if similar
quantities may be defined for massless quarks, given the fact that
mass singularities cancel in the variables we are considering, and thus they
remain 
well defined in the massless limit. This is not quite the case.
In fact, it is crucial for our discussion that the secondary heavy-flavour
production may be neglected. If this were not the case, evolution
of the fragmentation function would become multi-dimensional, involving
several flavour components and a singlet contribution.
Thus, the advantage of having a heavy flavour in the final state
lies in the fact that the evolution becomes essentially non-singlet,
so that we are left with a single component of the fragmentation
function, that simply cancels in the ratios we propose.

\section*{Acknowledgements}
We would like to thank P.~N.~Burrows and G.~Nesom for numerous discussions and
for making available to us experimental data on the $B$-$\overline{B}$
correlations.  Thanks also to P.~Uwer for discussions and comments on the
manuscript.  
The research of A.B. was supported by a Heisenberg grant of the DFG and
C.O. thanks the UK Particle Physics and Astronomy Research Council (PPARC) for
supporting his research.

\appendix
\section{Analytic results for $\boldsymbol{a_2,a_3,b_{22},b_{32},b_{33}}$}
\label{app:coefficients}
In this appendix, we collect the analytic results for the
coefficients $a_i$ and $b_{ij}$, $i,j=2,3$, defined in Eqs.~(\ref{eq:a_i})
and~(\ref{eq:b_ij}).   
Calling the generic coefficient
$F(z)$, where $z=m^2/s$ and $s=\ecm^2$, we write
\beq
F(z)={2C_F\over\beta\left[c_V(1+2z)+c_A(1-4z)\right]}\left[c_VF^V(z)
+c_AF^A(z)\right],
\eeq
where $\beta=\sqrt{1-4z}$ and $\omega=(1-\beta)/(1+\beta)$.
The electroweak couplings $c_V$ and $c_A$ are given by 
  \begin{eqnarray}
    \label{wcouplings}
    c_{V} &=&Q_b^2\, f^{\gamma\gamma} 
    + 2 \,g_V^b\,Q_b\, \Re\lq\chi(s)\rq\,f^{\gamma Z} 
    + g_V^{b\,2}\, |\chi(s)|^2\,f^{ZZ},\nonumber\\
    c_{A}  &=& g_A^{b\,2} |\chi(s)|^2 f^{ZZ},
\end{eqnarray}
with
\begin{eqnarray}
f^{\gamma\gamma}&=&1-\lambda_-\lambda_+ \,,\nonumber \\
f^{ZZ}&=&(1-\lambda_-\lambda_+)(g_V^{e2}+g_A^{e2})-
      2(\lambda_--\lambda_+) g_V^{e} g_A^{e}\,,\nonumber \\
f^{\gamma Z}&=&-(1-\lambda_-\lambda_+)g_V^e + 
      (\lambda_--\lambda_+)g_A^e\,.
\end{eqnarray}
Here, $Q_b=-1/3$ is the electric charge of the bottom quark, and
$g_{A/V}^{f}$ denote  the axial and vector couplings of the fermion 
$f$. In particular, $g_V^e = -\frac{1}{2} + 2 \sin^2\vartheta_W$, 
  $g_A^e =-\frac{1}{2}$ for an electron, and  
  $g_V^b = -\frac{1}{2} + \frac{2}{3} \sin^2\vartheta_W$,
  $g_A^b = -\frac{1}{2}$ for a bottom quark, where $\vartheta_W$ is the 
  weak mixing angle. The function
  $\chi(s)$ reads
  \begin{equation}
    \label{chi}
    \chi(s) = \frac{1}{4\sin^2\vartheta_W\cos^2\vartheta_W}\,
    \frac{s}{s-M_Z^2 + i M_Z \Gamma_Z},
  \end{equation}
  where $M_Z$ and $\Gamma_Z$ stand for the mass and 
  the width of the Z boson. 
  Finally,
  $\lambda_-$ ($\lambda_+$) denotes the longitudinal 
  polarization  of the electron (positron) beam.
\par
A simple calculation using the Born differential cross section for $e^+e^-
  \,\to  \, b \,\bar{b}\, g $ gives
\begin{eqnarray}
a_2^V(z)&=& 
\left(-{2\over 3}+z-{14\over 3}\*z^3\right)\*\ln(\omega)
+\left(-{11\over 3}+{7\over 6}\*z+7\*z^2\right)\*{\beta \over 3}
,\\
{a_2^A(z)-a_2^V(z)\over z}&=&
\left({8\over 3}-10\*z+12\*z^2+{20\over 3}\*z^3 \right)\*\ln(\omega)
+ \left({35\over 3}-{59\over 3}\*z-10\*z^2\right)
\*{\beta \over 3}
, \\
a_3^V(z)&=&\left(-{25\over 24}+{10\over 3}\*z
+{5\over 2}\*z^2-{32\over 3}\*z^3
-{11\over 12}\*z^4\right)\*\ln(\omega)
\nonumber \\ 
&+& \left( -{433\over 12}+{137\over 2}\*z
+{779\over 6}\*z^2+11\*z^3\right)\*{\beta \over 24}
,\\
{a_3^A(z)-a_3^V(z)\over z}&=&
\left({55\over 12}-20\*z+21\*z^2+{40\over 3}\*z^3
+{15\over 2}\*z^4\right)\*\ln(\omega)
\nonumber \\ 
&+& \left({823\over 36}-{845\over 18}\*z-{175\over 6}\*z^2
-15\*z^3\right)\*{\beta \over 4}
,\\
b_{22}^V(z) &=& 2\*z\*\(1+z^3\)\*\ln(\omega)
+\left({1\over 4}+{19\over 6}\*z-{1\over 6}\*z^2-z^3\right)\*\beta
, \\
{b_{22}^A(z)-b_{22}^V(z)\over z}&=& 
2\*z\*\(-2+3\*z-5\*z^3\)\*\ln(\omega)
+\left(-{3\over 4}-{17\over 6}\*z
+{5\over 6}\*z^2+5\*z^3\right)\*\beta
,\\
b_{32}^V(z) &=& 
z\*\left({19\over 6}+{1\over 3}\*z-2\*z^2+{17\over 3}\*z^3
+{1\over 3}\*z^4\right)\*\ln(\omega)
\nonumber \\ 
&+& \left({17\over 20}+{1721\over 180}\*z
+{47\over 45}\*z^2-{103\over 18}\*z^3-{1\over 3}\*z^4 \right)
\*{\beta \over 2}
,\\
{b_{32}^A(z)- b_{32}^V(z)\over z} &=&
z\*\left(-7+10\*z-10\*z^3-14\*z^4\right)\*\ln(\omega)
\nonumber \\ 
&+&\left(-{27\over 20}-{287\over 60}\*z
+{16\over 15}\*z^2+{37\over 6}\*z^3+7\*z^4\right)\*\beta
,\\
b_{33}^V(z) &=& 
z\*\left({14\over 3}-{13\over 6}\*z
-8\*z^2+{32\over 3}\*z^3+{16\over 3}\*z^4-z^5\right)\*\ln(\omega)
\nonumber \\
&+&\left({43\over 30}+{533\over 45}\*z
-{221\over 36}\*z^2-{1037\over 90}\*z^3
-{31\over 6}\*z^4+z^5\right)\*{\beta \over 2}
,\\
{b_{33}^A(z)-b_{33}^V(z)\over z} &=& 
z\*\left(-12+17\*z+{4\over 3}\*z^2
-12\*z^3-28\*z^4-{70\over 3}\*z^5\right)\*\ln(\omega)
\nonumber \\
&+&\left(-{109\over 15}-{661\over 30}\*z
+{53\over 20}\*z^2+{157\over 6}\*z^3+{287\over 6}\*z^4
+35\*z^5\right)\*{\beta \over 3}\,.
\end{eqnarray}
In the $m\,\to\,0$ limit, we have $\beta \,\to\, 1$, $\omega
\,\to\, z=m^2/s\,\to\,0$, so that
\begin{eqnarray}
a_2&=& 2\*C_F\*\left(-{2\over 3}\*\ln(z)-{11\over 9}\right)\,,\nonumber \\
a_3&=& 2\*C_F\*\left(-{25\over 24}\*\ln(z)-{433\over 288}\right)\,,\nonumber
\\
b_{22}&=& 2\*C_F\*{1\over 4}\,,\nonumber \\
b_{32}&=& 2\*C_F\*{17\over 40}\,,\nonumber \\
b_{33}&=& 2\*C_F\*{43\over 60}\,.
\end{eqnarray}
This behavior is well illustrated from the collections of values in
Table~\ref{tab:nlo_coeff}: while the $b_{ij}$ are finite in the $m\,\to\,0$
limit, the coefficients $a_i$ diverge as $\log(m/\ecm)$.

\section{Renormalization schemes}
\label{app:as_renorm}
Our calculation was carried out in the ``mixed''
renormalization  
scheme of Ref.~\cite{Collins:1978wz}, in which the light
flavours $n_{lf}$ are subtracted in the \MSB\ scheme, while the
heavy-flavour loops are subtracted at zero momentum.
In this scheme the heavy flavour decouples at low energy.
In fact, convergent heavy-flavour loops are suppressed by powers of the
mass of the heavy flavour. The only unsuppressed contributions come
from divergent graphs. But those are subtracted at zero external momenta,
so their contribution is removed by renormalization for small momenta.
In the mixed scheme, charge renormalization is given by the prescription
\beq
\label{eq:mixed}
\as^b =\mu^{2\ep} \tilde\as(\mu) \left\{ 1
- \tilde\as (\mu)
 \frac{1}{\bar\e} \left[ b_0^{(n_{lf})}
 - \left(\frac{\mu^2}{m^2}\right)^\ep
\frac{T_F}{3\pi} \right] + \ord{\as^2}\right\}\;,
\eeq
where $\as^b$ is the bare coupling constant, and $\tilde\as(\mu)$ is the
mixed-scheme 
coupling constant at the scale $\mu$. We have defined
\begin{equation}
b_0^{(n_{lf})}=\frac{11 C_A - 4 T_F n_{lf}}{12\pi}\,,
\end{equation}
and
\beq
\frac{1}{\bar\ep} = \frac{1}{\ep}+\log(4\pi)-\gamma_E\,,
\eeq
where dimensional regularization is used, with the number of space-time
dimensions equal to $4-2\ep$.

In the \MSB\ scheme the renormalization prescription is
\beq
\label{eq:MSbar}
\as^b = \mu^{2\ep} \as(\mu)  \left\{ 1
- \as(\mu) \frac{1}{\bar\ep} b_0^{(n_{f})} \right\}\;,
\eeq
where $n_f = n_{lf} + 1$, and $\as(\mu)$ is the \MSB\ coupling constant at the
scale $\mu$.
The couplings in the two schemes obey a 4-flavour and 5-flavour
renormalization-group equation respectively 
\begin{equation}
\frac{d}{d\log\mu^2} \tilde\as(\mu)=-b_0^{(n_{lf})}\tilde\as^2(\mu)\;,
\qquad\qquad
\frac{d}{d\log\mu^2} \as(\mu)=-b_0^{(n_{f})}\as^2(\mu)\;,
\end{equation}
that can be easily derived by imposing the constancy of the bare coupling
$\as^b$ under a renormalization group transformation.
Combining Eqs.~(\ref{eq:mixed})
and~(\ref{eq:MSbar}) we have
\beq
\label{eq:as_mixed_MSbar}
\tilde\as(\mu) = \as(\mu) - 
\frac{T_F}{3\pi}\, 
\log\(\frac{\mu^2}{m^2}\) \as^2(\mu) + \ord{\as^3}\,,
\eeq
which is the standard \MSB\ matching condition for flavour crossing.

In the heavy flavour production calculation we are considering, if
we express the result (usually given in terms of $\tilde\as$) in terms
of $\as$, an extra term arises, equal to
\begin{equation}
\label{extraTF}
- \mbox{Born} \times
\as(\mu)\frac{T_F}{3\pi}\, 
\log\(\frac{\mu^2}{m^2}\)\;.
\end{equation}
The only other diagrams where secondary heavy flavours appear
are the graphs with a gluon splitting into a heavy-flavour pair.
It can be shown that, if we neglect powers of $m/\ecm$, the
term in Eq.~(\ref{extraTF}) is exactly the term one needs to turn
the calculation of the heavy-flavour splitting, which is regulated
in the collinear limit by the heavy-flavour mass, into the corresponding
\MSB\ subtracted calculation, provided that interference terms of the
heavy flavours coming from gluon splitting with the primary heavy flavours
are fully neglected.

Equation~(\ref{extraTF}) follows immediately from the theory of heavy-flavour
fragmentation functions.
According to Ref.~\cite{Mele:1991cw}, Eq.~(3.16), the cross section
for the inclusive production of a heavy quark via gluon splitting,
with a fraction $x$ of the energy of the gluon, is
(at leading order and neglecting powers of $m/\ecm$)
\begin{equation}
\frac{d\sigma}{dx} = \sigma_0 \frac{\as T_F}{2\pi}\lq x^2+(1-x)^2\rq
\log\(\frac{\mu^2}{m^2}\) + \frac{d\hat\sigma}{dx}+{\cal O}(\as^2)\,,
\end{equation}
where $\sigma_0$ is the cross section for the production of the gluon
and $\hat\sigma$ stands for the massless
limit, \MSB\ subtracted cross section for the gluon splitting process.
Integrating both sides in $x$ one finds
\begin{equation}
\hat\sigma=\sigma-\sigma_0\frac{\as T_F}{3\pi}\log\(\frac{\mu^2}{m^2}\)\,.
\end{equation}
The second term on the right-hand-side is precisely the contribution of 
Eq.~(\ref{extraTF}).

\providecommand{\href}[2]{#2}\begingroup\raggedright\endgroup

\end{document}